\documentclass[12pt,a4paper]{article}
\usepackage[dvips]{graphicx}
\usepackage{times}
\usepackage[numbers]{natbib}
\newcommand\BibTeX{{\rmfamily B\kern-.05em \textsc{i\kern-.025em b}\kern-.08em
T\kern-.1667em\lower.7ex\hbox{E}\kern-.125emX}}

\usepackage{rawfonts}
\usepackage{graphicx,color,rotating,epsfig,multirow,multicol,hhline}
\usepackage{amsmath,amsthm,amssymb,amsfonts}

\usepackage{txfonts}    

\usepackage{graphics}

\usepackage{epsf}
\usepackage{latexsym}
\usepackage{bm}

\usepackage{amssymb}
\usepackage{latexsym}
\usepackage{amsmath}
\usepackage{multicol}
\usepackage{graphicx}

\newcommand{\defn}{\begin{quote}{\bf Definition. }}
\newcommand{\edefn}{\end{quote}}
\newcommand{\thm}{\begin{theorem}}
\newcommand{\ethm}{\end{theorem}}

\newcommand{\bp}{{\bm \beta}}
\newcommand{\bmat}[1]{\left [ \begin{array}{#1}}
\newcommand{\emat}{\end{array}\right ]}

\begin{document}

\title{A flexible multivariate random effects proportional odds model with application to adverse effects during radiation therapy}
\author{Nicole Augustin\textsuperscript{1}, Sung Won Kim\textsuperscript{2,3}\footnote{contributed equally to this work.},\\
Annemarie Uhlig\textsuperscript{4},  Christina Hanser\textsuperscript{4}, Michael Henke\textsuperscript{4}, Martin Schumacher\textsuperscript{2}\\
$^1$ Department of Mathematical Sciences, University of Bath, Bath, UK\\
$^2$Institute of Medical Biometry and Statistics, University of Freiburg\\
$^3$ Freiburg Center for Data Analysis and Modeling, University of Freiburg\\
$^4$ Section Clinical Studies, Department of Radiation Oncology,\\ University Hospital Freiburg}

\maketitle

\begin{abstract}

Radiation therapy in patients with head and neck cancer has a toxic effect on mucosa, the soft tissue in and around the mouth. Hence mucositis is a serious common side effect and is a condition characterized by pain and inflammation of the surface of the mucosa. Although the mucosa recovers during breaks of and following the radiotherapy course the recovery will depend on the type of tissue involved and on its location. Statistical models used in oncology for analysing oral mucositis are often too simplistic and there is a need for improved statistical methods in order to help improve planning of radiation therapy. We present a novel flexible multivariate random effects proportional odds model which takes account of the longitudinal course of oral mucositis at different mouth sites and of the radiation dosage (in terms of cumulative dose). The model is an extension of the {\em proportional odds model}  which is used for ordinal response variables. Our model includes the ordinal multivariate response of the mucositis score by location,  random intercepts for individuals and includes a non-linear function of cumulative radiation dose. The model allows to test whether sensitivity differs by mouth sites after having adjusted for site specific cumulative radiation dose. The model also allows to check whether and how the (non-linear) effect of site specific dose differs by site. 
We fit the model to longitudinal patient data from a prospective observation and find that after adjusting for cumulative dose, upper, lower lips and mouth floor are associated with the lowest mucositis scores and hard  and soft palate are associated with the highest mucositis scores. This implies the possibility that tissues at different mouth locations differ in their sensitivity to the toxic effect of radiation. We also find that cumulative dose followed by mouth site are the strongest predictors of mucositis, and the effects of age and gender are negligible.
\end{abstract}



\section{Introduction}
\label{sec : intro}
Radiation therapy in patients with head and neck cancer has a toxic effect on mucosa, the soft tissue in and around the mouth. Hence mucositis is a common side effect and is a condition characterized by pain and inflammation of the surface of the mucosa. 
Mucositis causes serious oral pain which makes the food intake of the cancer patients painful. 
It is known that mucositis is in general developed 14 days after starting radiotherapy. Although the mucosa recovers during breaks of and following the radiotherapy course the recovery will depend on the type of tissue involved and on its location. Hence there is a need to quantify the damage to the mucosa by different mouth sites. Here the clinical objective is in assessing the sensitivity to radiation of different mouth sites in order to improve planning of radiation therapy to avoid these side effects.

To obtain clinical evidence observational data were collected on a total of 75 patients receiving radiation therapy between September 2006 and February 2012. During the course of radiation therapy mucositis was assessed in 8 mouth sites approximately twice a week. For assessment an ordinal severity score with 5 levels was used, ranging from 0 (no change in tissue) to 4 (necrosis). At each assessment cumulative dose and the median value of the percentage site specific dose was recorded, so that the site specific cumulative dose could be derived. In addition all relevant patient characteristics were recorded. 
 
Statistical analyses on oral mucositis as an adverse effect of therapy are often  simplistic.  For example Sakellari et al. \cite{sakellari2015prospective} analyse oral mucositis in patients after high-dose chemotherapy using logistic regression of cumulative incidence of mucositis. The approach neglects the time course of the disease (oral mucositis), the fact the end-point is ordinal and the dosage of the chemotherapy. 

Our aim is to answer the clinical question by modelling the outcome variable mucositis score as a function of patient specific variables and (site) specific cumulative dose. Then the sensitivity of different mouth sites to radiation therapy can be assessed by estimating the effect of mouth sites on the mucositis score, after adjusting for confounder  variables, the most obvious being site specific cumulative dose. Due to the observational nature of the data we need to adjust for confounder variables. 

         These longitudinal data on the course of mucositis and cumulative dose of radiation therapy have a complex structure and pose a number of statistical challenges:

(1) the outcome is ordinal; (2) We have a multivariate outcome variable, as we have mucositis scores assessed for eight different sites per patient and time points (evaluation); (3) the outcome is longitudinal and we expect that repeated observations of individuals are correlated; (4) due to the fact that the mucosa can recover during radiation breaks a non-linear effect of sites specific cumulative dose on the mucositis score is plausible. (5) Finally, it is also likely that the effect of site specific dose differs by site, i.e. there is an interaction between site and site specific dose.  
 We address these challenges with a flexible multivariate random effects proportional odds model. This is an extension of the {\em proportional odds model} \cite{mccullagh1980regression} which is used for ordinal response variables and belongs to the class of generalized linear models used for modelling the dependence of an ordinal response on discrete or continuous covariates. The model is an extension of the multivariate random effects models for longitudinal data described in Verbeke et al. \cite{verbeke2014analysis}. See also  Faraway \cite{faraway2005extending} and Agresti \cite{agresti2014categorical}
for some recent expositions of ordinal response models. 


\section{The data}
\label{secdat}
Observational data were collected on a total of  75 patients receiving radiation therapy for head and neck cancer between September 2006 and February 2012 in the Radiation Oncology Department, Medical Center, University of Freiburg.  Three-dimensional radiation planning techniques and standard fractionation (5 x 2.0 Gy/wk) were followed. The total dose to the tumour was 60 Gy to 70 Gy. Radiation breaks were not allowed. Additional cisplatin 100 mg/m2 was given intravenously on days 1 and 22 (and on day 43 for patients undergoing primary definitive radiation or with residual cancer following surgery). The oral mucosa was evaluated in eight mouth sites (upper lip, lower lip, floor of mouth, right cheek, left cheek, tongue, soft palate hard palate) by using the five level NIH-CTC oral toxicity scale ((0) no mucositis, (1) erythema, (2) ulcer, (3) confluent ulcer, (4) necrosis). Twice-weekly assessments (at least 3 days apart) continued throughout radiotherapy and then until resolution of oral mucositis.
At each assessment the cumulative dose to the tumor and the median value of the percentage of the site specific dose was recorded, so that the site specific cumulative dose could be derived. In addition all relevant patient characteristics were recorded.
All study procedures were reviewed by the local ethics committee and all patients provided written consent.

\begin{figure}[!h]
 \centering
 \includegraphics[scale=0.7]{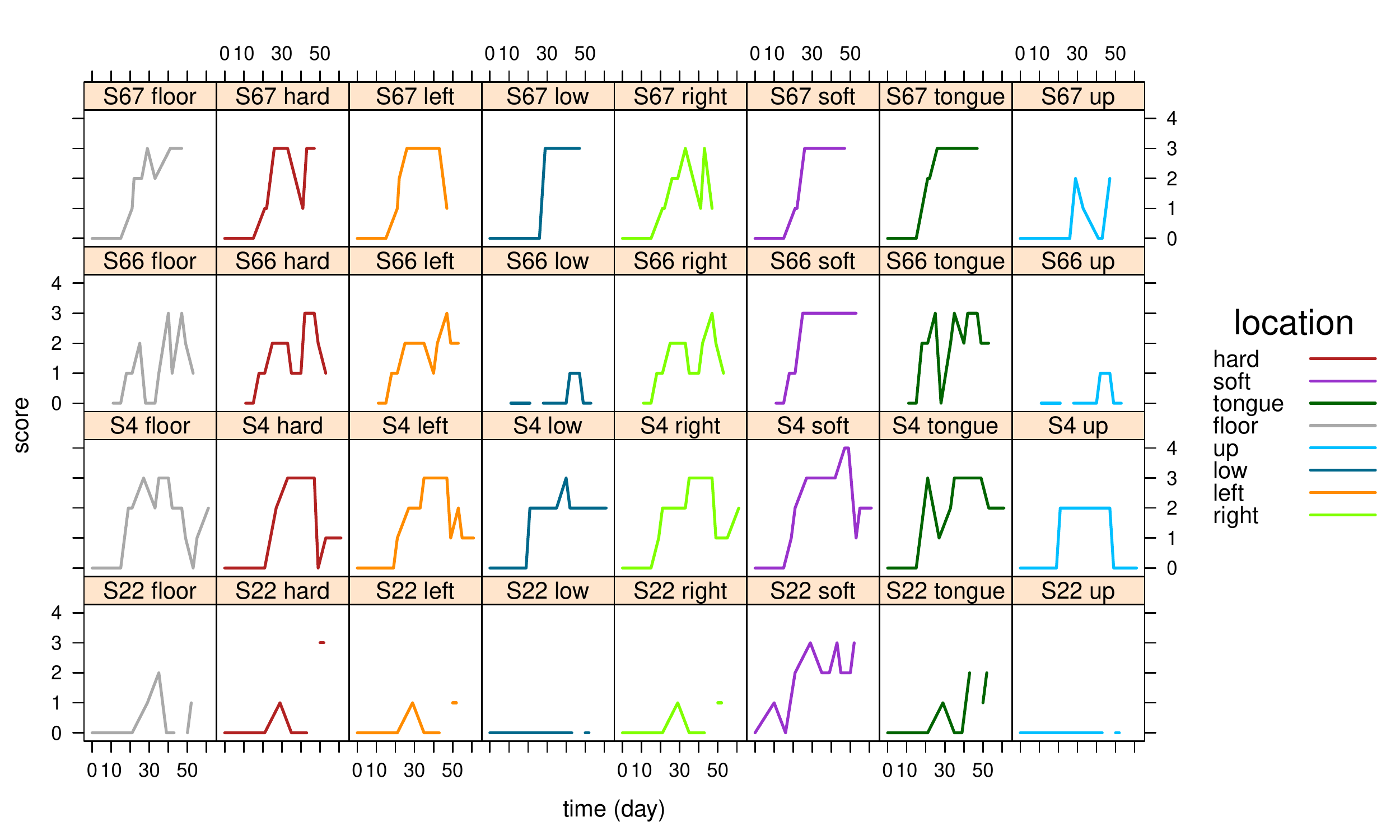} 
 \caption{Each row shows mucositis progression for one of four randomly selected individuals 
   on 8 mouth sites ({\tt hard} palate, {\tt soft} palate, {\tt tongue}, mouth {\tt floor}, 
{\tt upper} lip, {\tt lower} lip, left cheek, right cheek); the same color indicates the same site.}
 \label{fig:sample_mucositis}  
\end{figure}
\begin{table}[!h]
\begin{center}
\caption{Collapsing the four level ordinal mucositis scores into binary scores using different cut points $r$. Scores are no mucositis (0), erythema (1), ulcer (2), confluent ulcers (3) and necrosis (4). The total number of $n$ observations is due to 75 patients $\times$ 8 sites $\times$ 12.23 the mean number of evaluations per patient.}
\label{tab:collapse}
\begin{tabular}{|l|r|r|r|r|r|r|} \hline
&\multicolumn{5}{c|}{mucositis score}&\\\hline
cut point&0&1&2&3&4&$n$\\\hline
none&3641& 1616& 1433&  639&10&7339 \\\hline
$r = 0$&\multicolumn{1}{c|}{3641}&\multicolumn{4}{c|}{3698}&7339\\\hline
$r = 1$&\multicolumn{2}{c|}{5257}& \multicolumn{3}{c|}{2082}&7339\\\hline
$r = 2$ &\multicolumn{3}{c|}{6690}& \multicolumn{2}{c|}{649}&7339\\\hline
$r = 3$ &\multicolumn{4}{c|}{7329}& \multicolumn{1}{c|}{10}&7339\\\hline
 \end{tabular}
\end{center}
\end{table}
\begin{figure}[!h]
\centering
\includegraphics[scale=0.5]{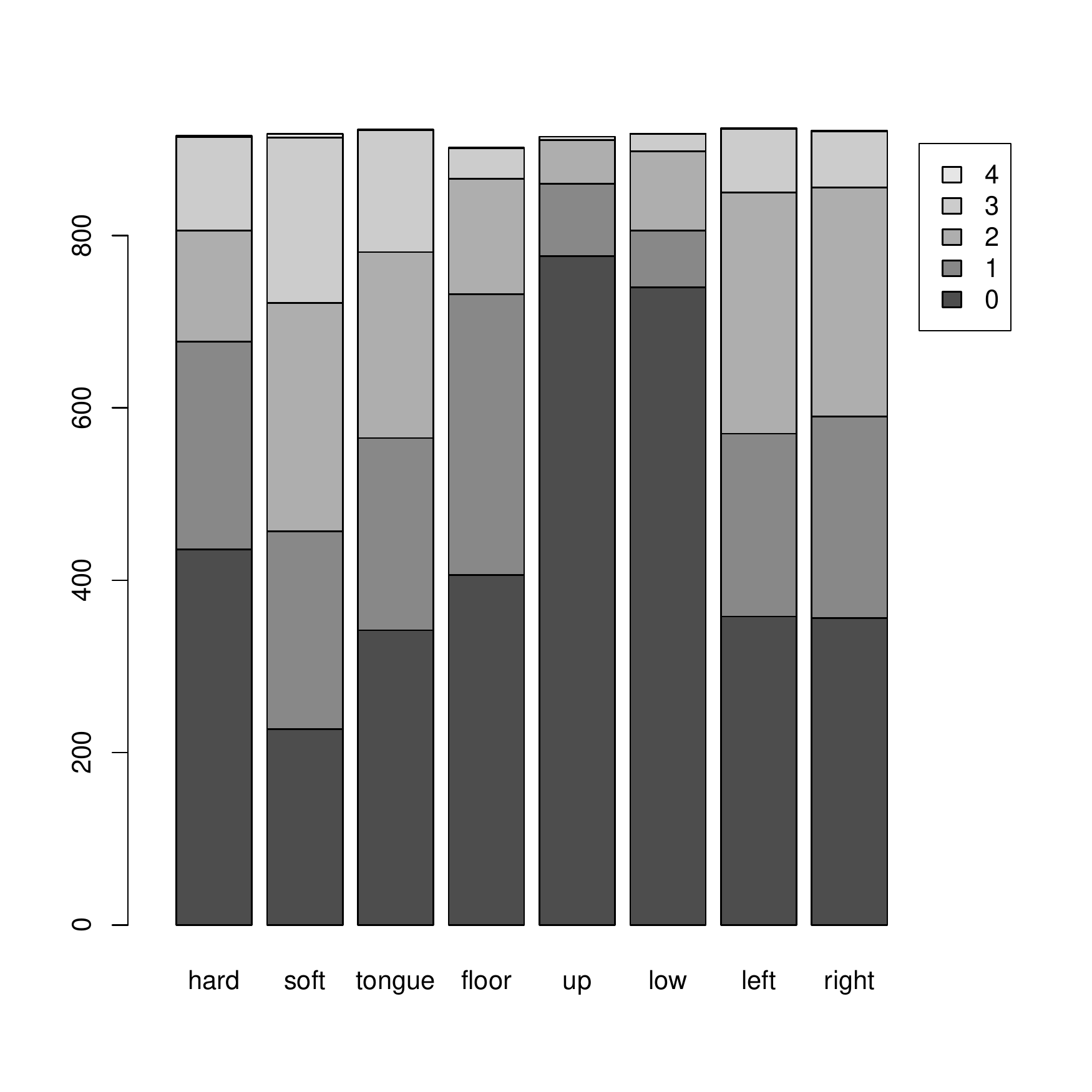}\includegraphics[scale=0.5]{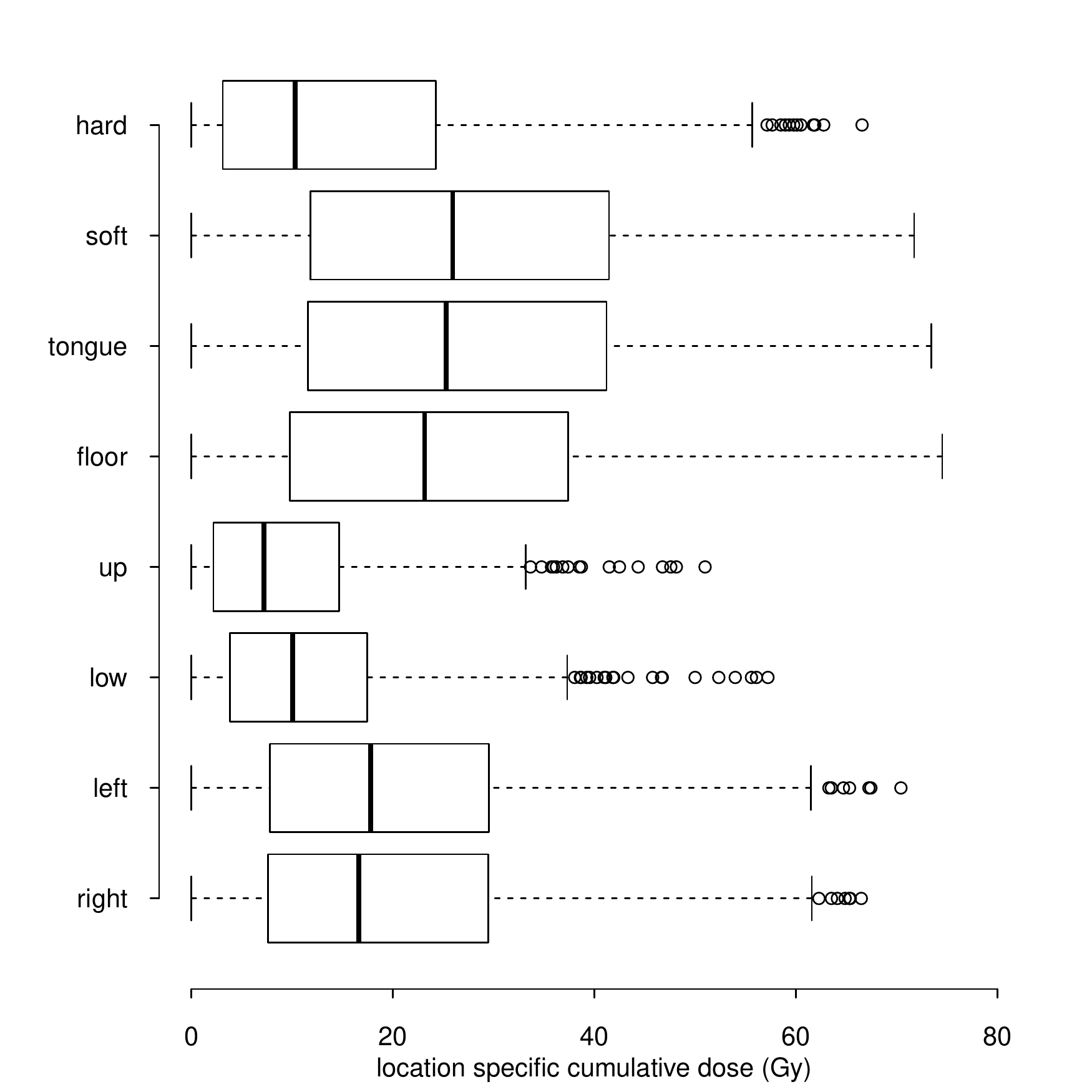} 
 \caption{Left: Frequencies of mucositis scores by site. Scores no mucositis (0), erythema (1), ulcer (2), confluent ulcers (3) and necrosis (4). Right: Boxplots of site specific median cumulative dose for mucositis score 0 to 4.}
 \label{fig:barplot}  
\end{figure}
%
\begin{figure}[!h]
\centering
\includegraphics[scale=0.9]{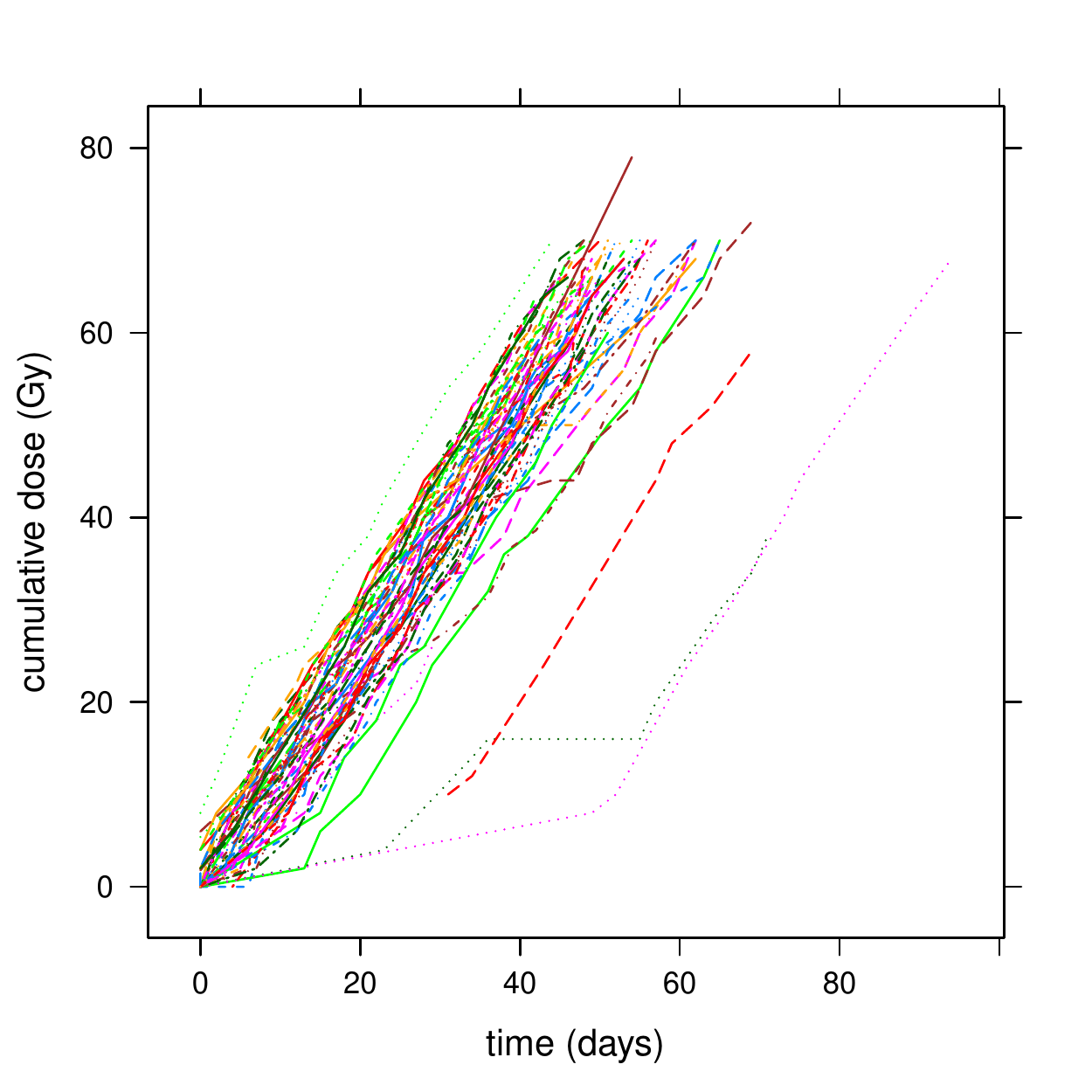} 
 \caption{Cumulative dose versus time in days.}
 \label{fig:cumdosl_studyt}  
\end{figure}

The mucositis progressions on eight different sites of four randomly sampled patients are depicted in Figure \ref{fig:sample_mucositis},
showing the ordinal scale of mucositis scores as well as longitudinal multivariate structure.  

Our main goal is to investigate the relationship between mucositis progression and potential risk factors. Let {\tt  y}$_{ijl}$ be the mucositis score for individual $i=1,\ldots, 75$, evaluation $j=1,\ldots, n_i$ and site $l=1,\ldots, 8$. This gives a total number of 7314 observations (75 patients $\times$ 8 sites $\times$ 12.23 the mean number of evaluations per patient). The risk factors are  patient characteristics ({\tt sex}$_i$ and {\tt age}$_i$), {\tt site}$_{l}$ (hard palate, soft palate, tongue, mouth floor, 
upper lip, lower lip, left cheek, right cheek), cumulative (radiation) dose ({\tt cumdose}$_{ij}$), 
median percentage dose of the site ({\tt perc}$_{il}$), the site specific cumulative dose {\tt cumdos.site}$_{ijl}$ and the site specific {\tt volume}$_{il}$. The site specific dose {\tt cumdos.site}$_{ijl}$ at evaluation $j$ is determined by multiplying the total cumulative dose {\tt cumdose}$_{ij}$ with  the median percentage dose of the site ({\tt perc}$_{il}$). 
The median percentage dose per site ({\tt perc}$_{il}$) is assumed to be constant over time as it is the median  percentage of {\tt cumdose}$_{ij}$ 
at the site receiving radiation over the entire radiation schedule.  
Apart for some exceptions, time in days is proportional to cumulative dose (Figure~\ref{fig:cumdosl_studyt}), hence in the following modelling we will use cumulative dose and site specific cumulative dose. 

Table~\ref{tab:collapse} shows that no mucositis (0) is observed most often, followed by erythema, ulcer, confluent ulcer and necrosis.
Figure~\ref{fig:barplot} shows frequencies of mucositis score by site and shows that on the lower and upper lip high mucositis scores are rarer than at other sites. 
Figure~\ref{fig:barplot} shows the {\tt cumdos.site}$_{ijl}$ by site and mucositis score: upper and lower lip tend to receive the lowest levels of site specific dose {\tt cumdos.site}$_{ijl}$ and mouth floor, tongue and soft palate tend to receive the highest doses.

\section{Biologically effective dose (BED)}

Here we analyse physically cumulative dose. To eventually correct for repopulation we also calculated the biologically effective dose (BED) using the so called $\alpha \beta$ model,  see for a review Fowler \cite{fowler2010}. 
Here we assume $\alpha \over \beta$	 = 10Gy, where $\alpha$ and $\beta$ are the coefficients for dose and dose$^2$ in a linear model for cell kill. In the case of mucosa the $\alpha$	= 0.3Gy$^{-1}$, the onset of accelerated repopulation is 7 days and the average doubling time during accelerated repopulation is assumed to be 2.5.
 
\begin{figure}[!htb]
 \centering
 \includegraphics[scale=0.8]{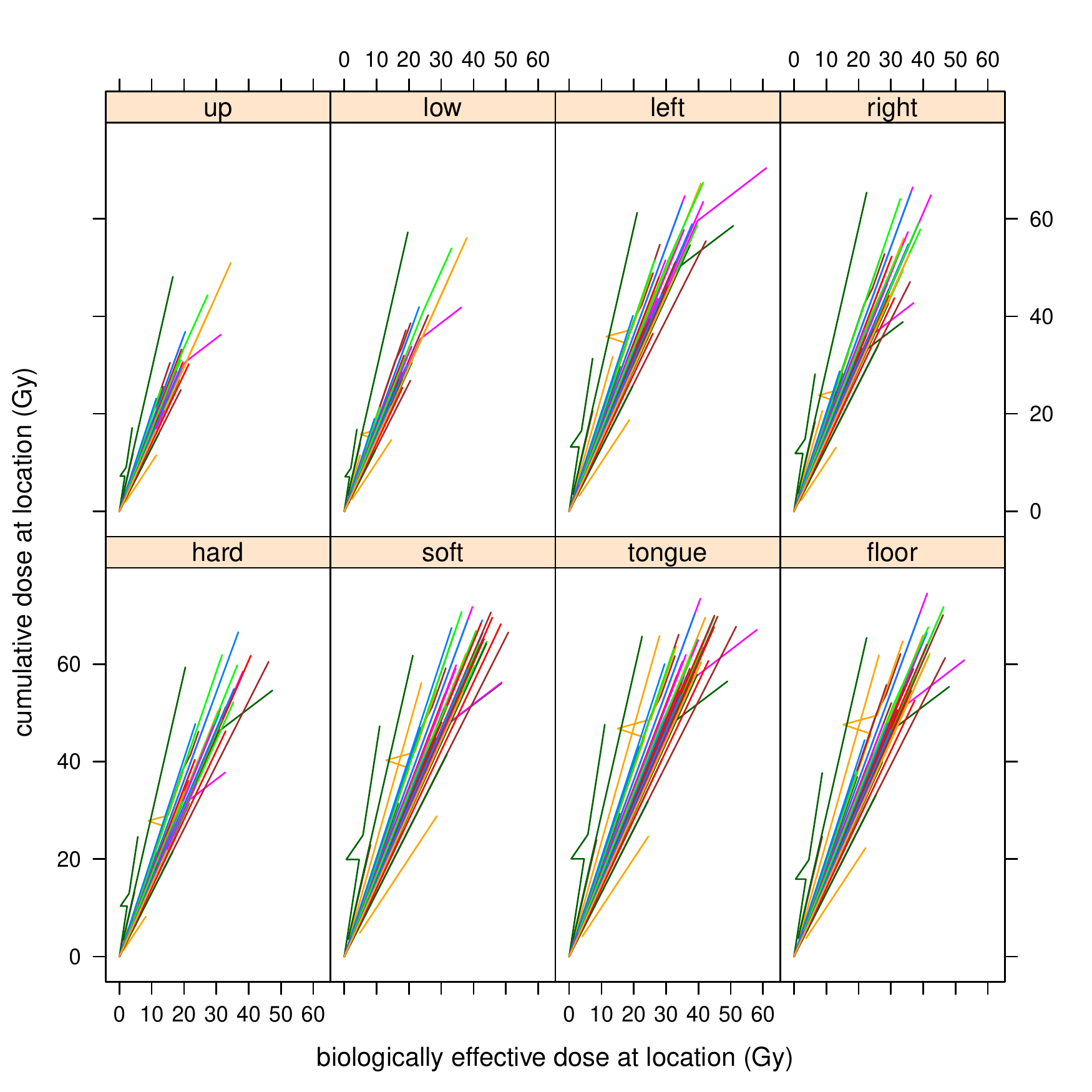}
 \caption{Cumulative dose vs. acute mucosa biological effective dose (BED).} 
 \label{fig:cumdosl_BEDl}
\end{figure}

Figure \ref{fig:cumdosl_BEDl} compares BED versus cumulative dose for each site. The straight lines, except for some kinks, indicate that the two measures are proportional to each other. The kinks originate from breaks in the radiation schedule due to strong side effects for some patients. These breaks affect BED but not the cumulative dose. 

\section{The flexible proportional odds model with random effects}
\label{sec : meth}
As outlined in the introduction due to the complex data situation we require a model for an ordinal multivariate outcome variable.  In addition the model should be able to accommodate the correlated repeated measures of individuals and should allow effects of continuous variables to be non-linear, and we also want to test  whether there is an interaction between site and site specific dose.  

\subsection{The proportional odds model} The proportional odds model \cite{mccullagh1980regression} addresses the first point, as it is a class of generalized linear models used for modelling the dependence of an ordinal response on discrete or continuous covariates. It can also be seen as an extension of a logistic regression model  \cite{bender1997ordinal}, as we could model the mucositis scores collapsed into two categories for the two events $Y \leq r$ and $Y > r$ as shown in Table~\ref{tab:collapse}. But this leads to loss of efficiency, resulting in larger standard errors \cite{agresti2014categorical}.

Let $Y_i$ denote the mucositis severity score in the range 0, ..., 4 for individual $i$. Then we model the cumulative probability of $Y_i$ being less or equal to mucositis score $r$, $\text{P}(Y_i \leq r)$ on the logit scale. That is on the log scale of the (cumulative) odds of the event $Y_i \leq r$:
\begin{equation}
\text{logit}(\text{P}(Y_i \leq r)) = \log\frac{\text{P}(Y_i\leq r)}{\text{P}(Y_{i} > r)}  ={\alpha}_r-x_i\beta.
\label{mod.logit}
\end{equation}
On the logit scale the model is linear which makes it computationally convenient. In the above example the $x_i$ is a continuous variable, e.g. cumulative dose. 
In the case of mucositis a plausible way of constructing model \ref{mod.logit} is to use the concept of an unobserved continuous response. Suppose $U_i$ is the unobserved mucositis severity on a continuous scale with $U_i = {x}_{i}{\beta} +\epsilon_i$.  Although we only observe the discrete mucositis scores, $Y_i = r$ is only observed if $\alpha_{r-1} < U_i \leq \alpha_r$. The "cut-points" $\alpha_r$ are unknown points on the continuous mucositis severity scale. Lets also assume that $\epsilon_i = U_i - x_{i}{\beta}$ has a logistic distribution function $F(.)$ with $F(.) = \frac {\text{exp}(.)} {1 + \text{exp}(.)}$ then: 
 $$\text{P}(Y_i\leq r)=\text{P}(U_i \leq \alpha_r)=\text{P}(U_i - {x}_{i}{\beta} \leq \alpha_r - {x}_{i}{ \beta}) = F(\alpha_r - {x}_{i}{\beta})$$  
 and 
 \begin{equation*}
 \text{P}(Y_i \leq r) = \frac {\exp ({ \alpha}_r-{x}_{i}{\beta})} {1 + \exp ({\alpha}_r-{x}_{i}{\beta})}
 \label{mod.probscale}
\end{equation*}
This model is mathematically equivalent to model \ref{mod.logit}. We also note the intuitive interpretation, if $\beta > 0$ as $x_i$ increases also the $\text{P}(Y_i > r)$ increases. For example, as cumulative dose increases, it is more likely that a high mucositis score is observed. 
In a model containing  a linear effect for {\tt cumdos.site} the odds ratio of the event $Y \leq r$ for $x_1 = \text{cumdos.site} - 10 $ and $x_2 = \text{cumdos.site}$, whilst keeping all other variables fixed, is 
 \begin{equation*}
  \frac{\text{P}(Y_i\leq r|x_1)/\text{P}(Y_i> r| x_1)}
  {\text{P}(Y_i\leq r|x_2)/\text{P}(Y_i> r|x_2)}
  =\exp(- (x_1 - x_2) \beta)= \exp( (x_2 - x_1) \beta) = \exp( 10 ~\beta). 
 \end{equation*}
When using the same $\beta$ across all categories, we assume that the odds ratio does not depend on the category $r$, i.e. these cumulative odds are proportional across all categories. In a model containing a factor variable for site based on the frequencies in Figure~\ref{fig:barplot}, this implies that the cumulative odds of the events $Y \leq r$, i.e. $Y \leq 0$, $Y \leq 1$, $Y \leq 2$ and $Y \leq 3$ by site are proportional across all categories. For example, comparing the cumulative odds of mucositis in hard palate versus soft palate is the same for all $r$:
\begin{equation*}
  \frac{\text{P}(Y_i\leq 0|\text{hard})/\text{P}(Y_i> 0| \text{hard})}
  {\text{P}(Y_i\leq 0|\text{soft})/\text{P}(Y_i> 0|\text{soft})}
  = ...= \frac{\text{P}(Y_i\leq 2|\text{hard})/\text{P}(Y_i> 2| \text{hard})}
  {\text{P}(Y_i\leq 2|\text{soft})/\text{P}(Y_i> 2|\text{soft})} 
 \end{equation*}

\subsection{A flexible multivariate random effects proportional odds model}
In order to address all of the model requirements we extend the proportional odds model by incorporating a multivariate response, random effects and non-linear functions of continuous variables. This  is an extension of the class of generalised additive mixed models \cite{fahrmeir2013regression,Wood06}, because the logistic distribution we are assuming for the latent response variable $U_i$ and the error $\epsilon_i$ does not belong the exponential family. Below we introduce these different features step by step.

The model for individual $i$, assessment time index $j$, and site $l$
is specified as 
\begin{equation*}
 \text{logit P}(Y_{ijl}\leq r)=\log\frac{\text{P}(Y_{ijl}\leq r)}{\text{P}(Y_{ijl} > r)}=\alpha_r - \eta_{ijl}
\end{equation*}
where 
\begin{equation}
\label{eqn:linear}
\mbox{linear model:~~~~~} \eta_{ijl} = {\bf x}_{ijl}^T \bp + b_i
\end{equation}
and $U_{ijl}= \eta_{ijl}+\epsilon_{ijl}$ and $\epsilon_{ijl}$ follows  the logistic distribution. The ${\bf x}_{ijl}$ is a row of the model matrix containing the explanatory variables {\tt age}$_i$, {\tt study}$_i$, {\tt sex}$_i$, {\tt cumdos.site}$_{ijl}$, {\tt site}$_i$, where {\tt site}$_i$ is a vector with dummy variables for the 8 sites and the parameter vector $\bp$ contains the respective coefficients.  We deal with the multivariate response, the mucositis scores assessed for eight different mouth sites per patient and evaluation by including the factor {\tt site}$_i$. This allows us to test the null hypothesis that the effects of site are equal.  As the dose specific to site is determined by the tumour location, we adjust site estimates by site specific dose {\tt cumdos.site}$_{ijl}$. 
Individuals are observed repeatedly in time and within individual observations $y_{ijl}$  are correlated  in time and between locations and we account for this by introducing a random intercept $b_i$ for individual $i$ with $b_{i}\sim {N}(0, \sigma^2_b)$.
As we expect the dose-response relationship not to be linear, we use a flexible function to model the effect of sites specific cumulative dose:
\begin{equation}
\label{eqn:final}
\mbox{reduced model:~~~~~} \eta_{ijl} = {\bf x}_{ijl}^T \bp + f(\text{cumdos.site}_{ijl})+ b_i 
\end{equation}
the smooth function $f$ is represented using a thin plate regression spline \cite{Wood06} with a penalty based on the second derivative of the smooth function and now ${\bf x}_{ijl}^T =$ ({\tt age}$_i$, {\tt study}$_i$, {\tt sex}$_i$, {\tt site}$_i^T$). 
 The model above can be further extended. Besides testing whether sites effects differ, we are also interested in testing whether the effect of site specific dose differs by site. For this we  introduce an interaction between site and site specific dose, by letting $f$ vary by {\tt site}$_l$:
\begin{equation}
\label{eqn:interaction}
\mbox{full model:~~~~~} \eta_{ijl} = {\bf x}_{ijl}^T \bp + f_l(\text{cumdos.site}_{ijl}) + b_i .
\end{equation}

\subsection{Parameter estimation}

Inference can be based on Markov chain Monte Carlo (MCMC) \cite{FaKn204, augustin2007spatial}, on a mixed model approach  using restricted maximum likelihood (REML) \cite{kneib2006structured, fahrmeir2013regression,  WoodPyaSae2015}. All of these are implemented in the BayesX package \cite{Belitz2015, Umlauf2015}. We present estimates using a Laplace approximation to REML of Wood et al. \cite{WoodPyaSae2015}, implemented in the {\tt gam()} function of the  {\tt R} (\cite{Rman14}) package {\tt mgcv}. 
In the supplementary material we supply example R code for model fitting in both {\tt BayesX} and in {\tt mgcv}.

Although the above reduced model~(\ref{eqn:final}) and the full model (\ref{eqn:interaction}) include additive non-linear  predictors and random effects they can be expressed as a penalised generalised linear model (GLM);
$ \eta_{ijl} = {\bf x}_{ijl} {\bm \theta}$, 
 where ${\bf x}_{ijl}$ is a row of the model matrix containing {\em all} components of the model; that is, all strictly parametric components, such as explanatory variables of linear effects and the basis functions for non-linear predictors evaluated at observations $ijl$. The parameter vector $\bm \theta$ contains the coefficients of the basis functions and coefficients of linear terms and the cut-points $\alpha_1$ to $\alpha_4$ . Note that for identifiability the first cut point $\alpha_1$ is set to -1.

 The parameter estimation is an extended version of the nested iteration scheme described in Wood \cite{Wood2011}. Due to the ordered categorical response the scheme also needs to accommodate the  variable number of ordered cut points (here $\alpha_1$ to $\alpha_4$).  The outer iteration is a Laplace approximation marginal (restricted) maximum likelihood (LAML) estimation of smoothness parameters and the cut points, where the coefficients of linear terms, and of basis functions for the smooth terms of the parameter vector  ${\bm \theta}$, are integrated out.   
The inner iteration is a penalised iterative re-weighted least squares (PIRLS) algorithm to find all other parameters, i.e. the coefficients of basis functions, and coefficients of linear terms. For details see Appendix F in Wood et al. \cite{WoodPyaSae2015}.

\subsection{Model selection and validation} 
\label{mod:sel}

We consider the Akaike information criterion (AIC) and the Bayesian information criterion (BIC)for model selection. The AIC is defined as  $AIC = - 2 {\rm l}(\hat {\bm \theta}) + 2 edf$, the BIC is $BIC = - 2 {\rm l}(\hat {\bm \theta}) + \log(n) edf$, where ${\rm l}$ is the conditional log likelihood given the penalised parameters, $edf$ are the effective degrees of freedom, here estimated by the trace of the matrix that maps the un-penalised estimates onto the penalised estimates corrected for the  smoothing parameter uncertainty as described in Wood et al. \cite{WoodPyaSae2015}; see also Greven and Kneib \cite{greven2010}. 
We also consider the root mean square prediction error (RMSPE) estimated by cross-validation as a selection criterion.  For the cross-validation we leave out in turn each of 15 mutually exclusive sets of patients and fit the model to the remaining patients data and predict the probabilities for the outcomes, the mucositis scores. Having obtained predictions for each set of patients left out we can estimate the root mean prediction error for each mucositis score.

\section{Results}

\begin{figure}[!htb]
 \centering
\includegraphics[scale=0.6]{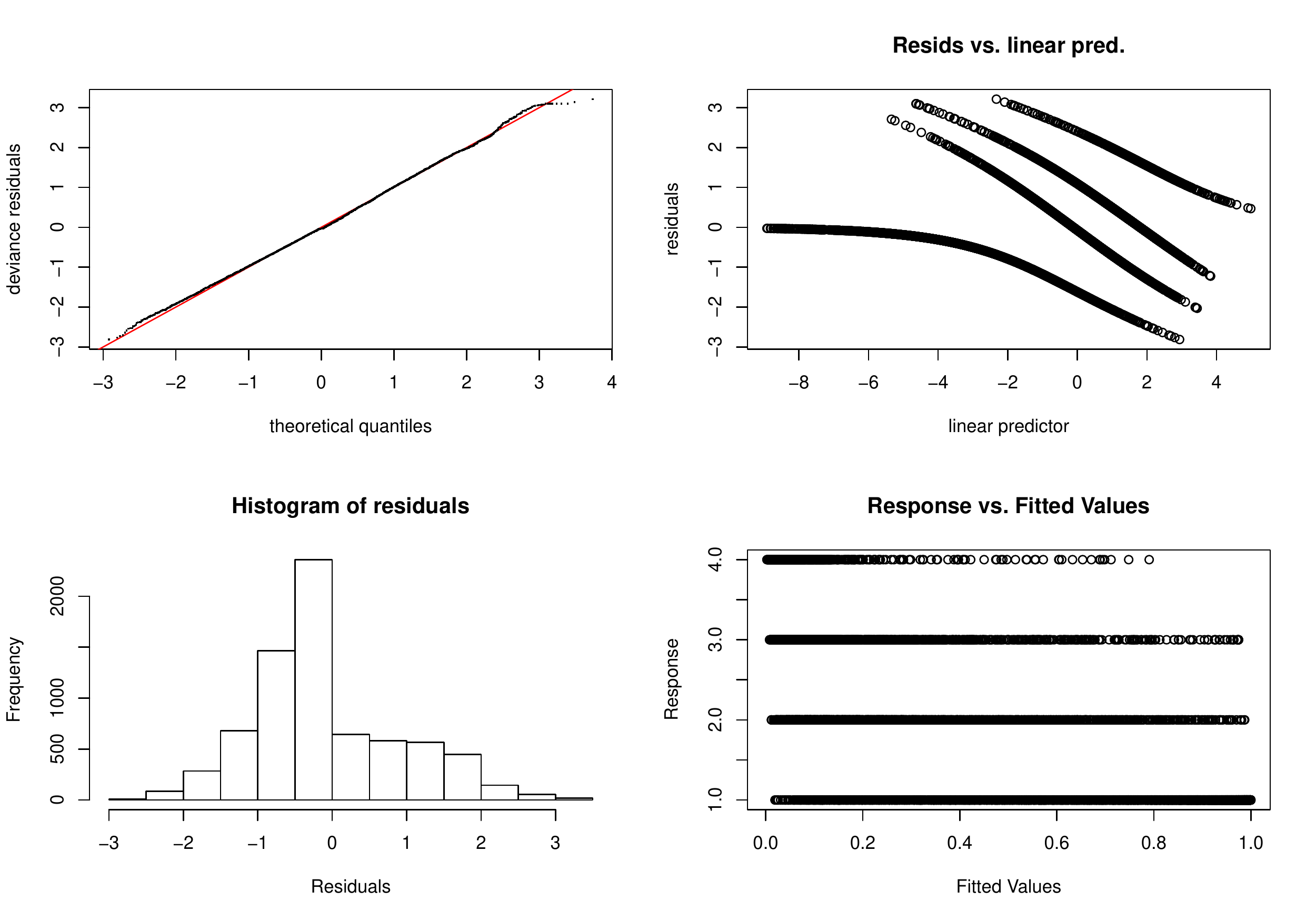}
 \caption{Residual plots of the best model (\ref{eqn:best}).}
 \label{resi:fig}
\end{figure}

\begin{figure}[!htb]
 \centering
 \includegraphics[scale=0.6]{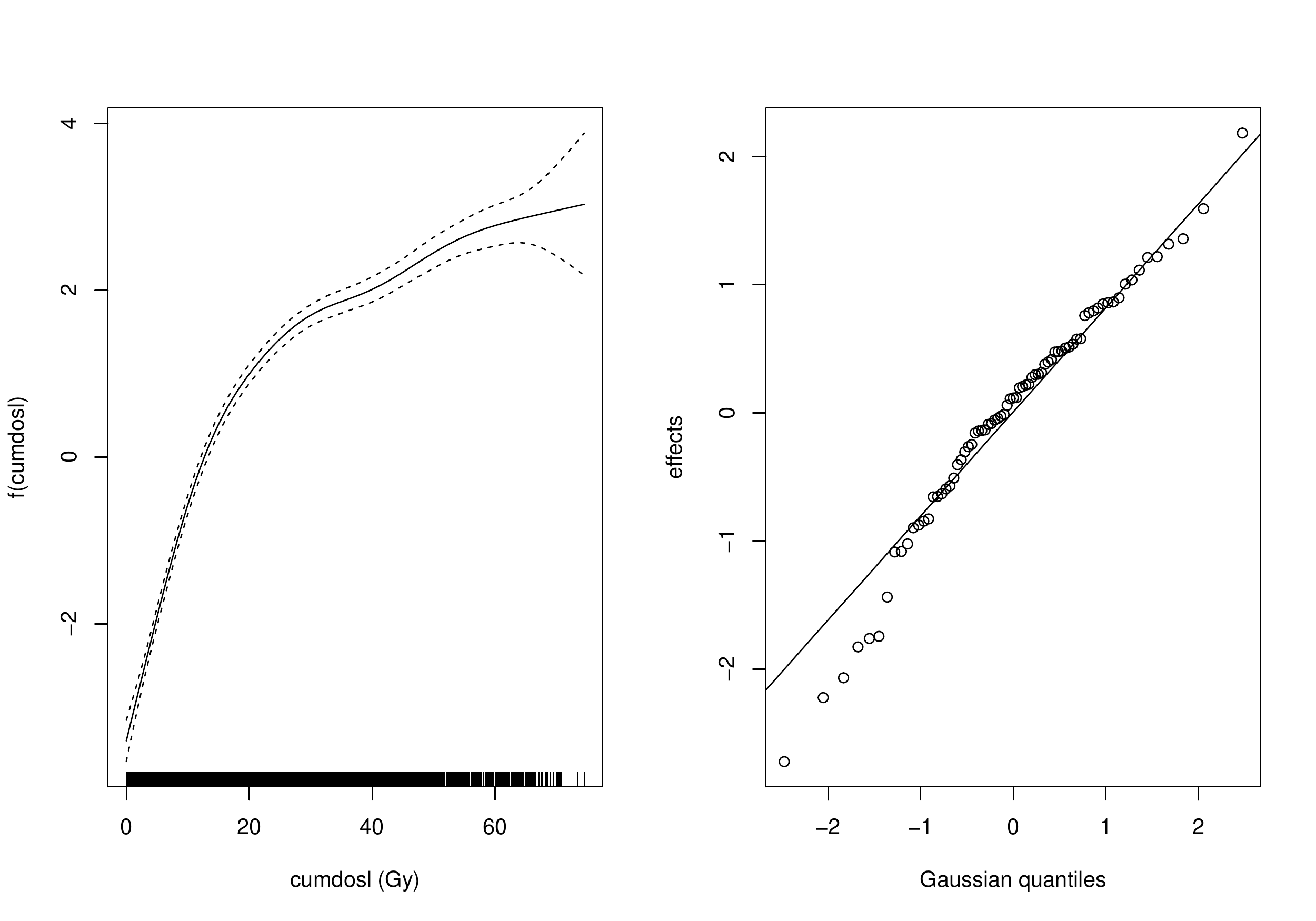}
 \caption{Best model (\ref{eqn:best}): Smooth (left) and qq-plot of the random intercepts for individuals (right).}
 \label{Smooth:fig}
\end{figure}

In the modelling we combine the mucositis scores 3 (confluent ulcer) and 4 (necrosis) due to the low frequencies of score 4. We start the model selection by reducing the complexity of the full model~(\ref{eqn:interaction}). Table~\ref{mod:tab} gives an overview of the considered models and the results of selection criteria are shown in Table~\ref{modsel:tab}. The most complex model is the full model~(\ref{eqn:interaction}) overall has marginally the best goodness of fit, but with regards to parsimony the reduced model~(\ref{eqn:final}) with the terms for age and sex removed comes out best. Hence the final selected model is
\begin{equation}
\label{eqn:best}
\mbox{best model:~~~~~} \eta_{ijl} =  \alpha +\beta_1 \tt{study}_{i} + \tt{site}_{i}^T \bp_2 + f(\tt{cumdos.site}_{ijl})+ b_i 
\end{equation}
where $\bp_2$ is the parameter vector for the 8 sites. 
Checking the residuals of the best model~(\ref{eqn:best}), shows that the model fits well. The distributional assumptions of the model are correct, i.e. the residuals follow a logistic distribution (Figure~\ref{resi:fig})  and the random effects follow a normal distribution (Figure~\ref{Smooth:fig}). The assumption of proportional odds is adequate and has been checked using diagnostic plots (not shown). 

\begin{figure}[!htb]
 \centering
 \includegraphics[scale=0.6]{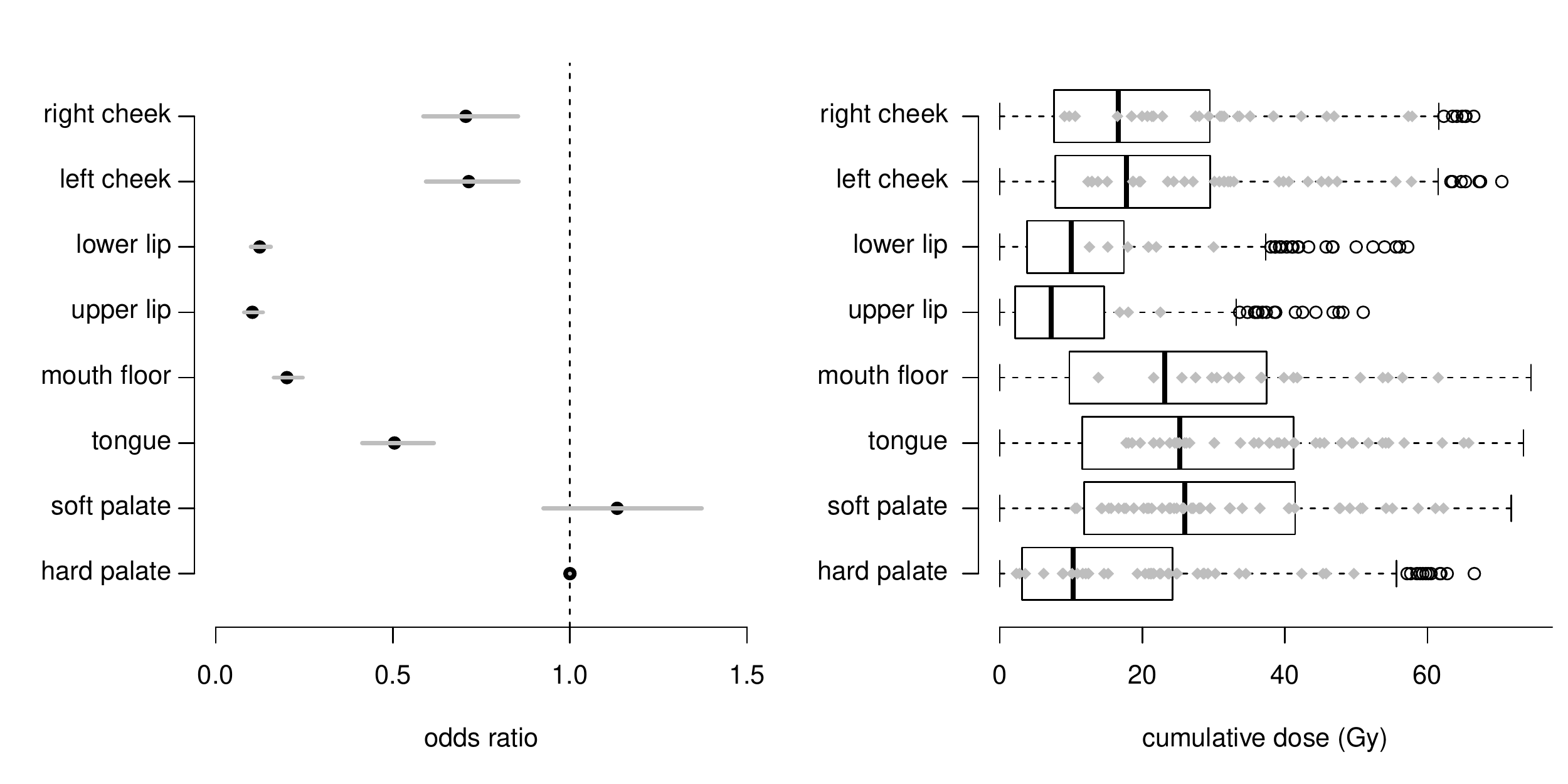}
 \caption{Best model (\ref{eqn:best}). Odds ratios of all the sites with 95\% confidence/credible intervals (left) and corresponding box-plots of cumulative doses applied to the eight sites (right). They 
 They are marked with gray dots by the cumulative doses when mucositis 3 or 4 first occurred.}
 \label{Oddsfirstobs:fig}
\end{figure}

The model selection results also show that {\tt cumdos.site} is by far the most important predictor, followed by {\tt site}. Results also show that effect of site specific cumulative dose is not linear, as the goodness of fit is substantially reduced when the we use a linear effect for {\tt cumdos.site}.  This is also apparent from the estimated smooth function of {\tt cumdos.site} shown in Figure~\ref{Smooth:fig}. Replacing {\tt cumdos.site} by {\tt BED} in the full model~(\ref{eqn:interaction}), the reduced model~(\ref{eqn:final}) and the linear model~(\ref{eqn:linear}) does not improve the model fit. These models give an AIC of  15028.67, 15079.57 and 15085.05 respectively and the BIC values are 15726.93, 15658.66 and 15639.53.
The effect of {\tt cumdos.site} is shown on the scale of the unobserved continuous mucositis score. The cumulative dose has initially, up to approximately 20~Gy  quite a strong effect on mucositis, and this effect becomes weaker for $>$ 20 Gy. There is a considerable amount of uncertainty about the effect for high cumulative dosages. Figure~\ref{Oddsfirstobs:fig} shows the cumulative odds of having a score greater than $r$ over having a score less or equal to $r$, with hard palate as the reference. e.g.: 
\begin{equation*}
 \frac{\text{P}(Y_i > r|\text{soft})/\text{P}(Y_i \leq r| \text{soft})}
  {\text{P}(Y_i> r|\text{hard})/\text{P}(Y_i\leq r|\text{hard})} =\frac{\text{P}(Y_i \leq r| \text{hard})/\text{P}(Y_i > r|\text{hard})}
  {\text{P}(Y_i\leq r|\text{soft})/\text{P}(Y_i > r|\text{soft})} = exp(0.185)= 1.171.
   \end{equation*}
This shows that, after adjustment for cumulative radiation dose,  upper, lower lip and mouth floor are much less likely to have a higher mucositis score than the other sites. The site soft palate is most likely to have a high score.

\begin{table}[ht]
\label{mod:tab}
\centering
\caption{Model overview given in ascending order regarding model complexity. The model equation numbers are given in brackets. The ${\bf x}_{ijl}$ is a row of the model matrix containing the explanatory variables {\tt age}$_i$, {\tt study}$_i$, {\tt sex}$_i$, {\tt cumdos.site}$_{ijl}$, {\tt site}$_i$, where {\tt site}$_i$ is a vector with dummy variables for the 8 sites and the parameter vector $\bp$ contains the respective coefficients.
}

  \begin{tabular}{ll}
&\\ 
   \hline
model  & formula\\ 
   \hline 
   linear model (\ref{eqn:linear})  &$\eta_{ijl} = {\bf x}_{ijl}^T \bp + b_i$\\
 best model (\ref{eqn:best}) & $\eta_{ijl} =  \alpha + \beta_1 \tt{study}_{i} + \tt{site}_{i}^T \bp_2 + f(\tt{cumdos.site}_{ijl})+ b_i$ \\
 reduced model (\ref{eqn:final}) &$\eta_{ijl} = {\bf x}_{ijl}^T \bp + f(\text{cumdos.site}_{ijl})+ b_i $\\

  full model (\ref{eqn:interaction})&$\eta_{ijl} = {\bf x}_{ijl}^T \bp + f_l(\text{cumdos.site}_{ijl}) + b_i$ \\
   \hline
\end{tabular}
\end{table}

\tiny
\begin{table}[ht]
\label{modsel:tab}
\centering
\caption{Model selection results. Shown are the effective degrees of freedom (edf), Akaike information criterion (AIC), Bayes information criterion (BIC), root mean squared error for predicting mucositis score 0 (RPE0), ... , 3 (RPE3)  on the probability scale.}
  \begin{tabular}{rrrrrrrrr}
  &&&&&&&&\\
   \hline
  & N & edf & AIC & BIC & RPE0 & RPE1 & RPE2 & RPE3 \\ 
   \hline
  full model (\ref{eqn:interaction})  & 7339 & 121.92 & 12185.50 & 13085.60 & 0.367 & 0.397 & 0.373 & 0.265 \\ 
reduced model (\ref{eqn:final})  & 7339 & 84.71 & 12380.21 & 12989.66 & 0.371 & 0.398 & 0.374 & 0.265 \\ 
linear model (\ref{eqn:linear})   & 7339 & 78.38 & 13076.64 & 13639.67 & 0.380 & 0.401 & 0.375 & 0.272 \\ 
linear model (\ref{eqn:linear})&&&&&&&& \\ 
- {\tt cumdos.site}   & 7339 & 77.76 & 15280.12 & 15835.42 & 0.455 & 0.407 & 0.390 & 0.278 \\ 
reduced model (\ref{eqn:final})~~- {\tt site}    & 7339 & 76.71 & 13284.12 & 13836.23 & 0.392 & 0.404 & 0.377 & 0.269 \\ 
reduced model (\ref{eqn:final})~~- {\tt study}   & 7339 & 85.05 & 12380.37 & 12992.07 & 0.373 & 0.398 & 0.374 & 0.267 \\ 
reduced model (\ref{eqn:final})~~- {\tt age}   & 7339 & 84.60 & 12380.17 & 12988.85 & 0.371 & 0.398 & 0.374 & 0.265 \\ 
reduced model (\ref{eqn:final})~~- {\tt sex}   & 7339 & 84.60 & 12380.17 & 12988.86 & 0.370 & 0.398 & 0.374 & 0.265 \\ 
reduced model (\ref{eqn:final})&&&&&&& \\ 
- {\tt sex} - {\tt age}  & 7339 & 84.48 & 12380.13 & 12988.03 & 0.370 & 0.398 & 0.374 & 0.265 \\ 

 reduced model (\ref{eqn:final}) &&&&&&&&\\
 - {\tt sex} - {\tt age} - {\tt study}  & 7339 & 84.85 & 12380.29 & 12990.63 & 0.372 & 0.397 & 0.373 & 0.266 \\ 
   \hline
\end{tabular}
\end{table}

\normalsize

\clearpage
\section{Discussion}
\label{sec:discuss}
In this observational study the radiotherapists made an effort to prevent much damage of mucositis and randomized assignment of radiation dose for each site was not feasible. The assigned dose has been carefully planned to avoid 
severe mucositis especially for what the practitioners thought are the most vulnerable sites, upper and lower lip. 
Figure~\ref{fig:barplot} confirms that the distribution of cumulative dose is not balanced among sites. In particular lower and upper lip  and hard palate obtain the lowest median cumulative doses of 7.2 to 10.3 Gy. The highest median doses of 23.1 to 25.9 Gy arrive at soft palate, tongue and mouth floor. The left and right cheek have median doses of 16.6 to 17.8 Gy. When we investigate the frequencies of the higher non-zero mucositis scores by site in Figure~\ref{fig:barplot}, the same order is reflected. Upper and lower lip and hard palate have the lowest frequencies of mucositis scores and soft palate receives the highest frequencies of mucositis scores. 
If the dose was assigned to each site for each individual randomly, 
it would have been possible to estimate the site effect without adjusting for (cumulative) dose. 
However, the assigned dose has been carefully planned to avoid 
severe mucositis especially for the most vulnerable sites, that is the sensitivity of sites and site specific cumulative dose are confounded. For estimating the sensitivity of sites we need to adjust for cumulative dose in a statistical model. 
The selected best model~(\ref{eqn:best}) contains site specific cumulative dose as a non-linear effect, and the results show that after adjusting for site specific cumulative dose we find that sensitivity to radiation varies by site.
The upper and lower lip are, after adjusting for cumulative dose, the least vulnerable to radiation, followed by mouth floor (Figure~\ref{Oddsfirstobs:fig}). More vulnerable than the latter are the sites tongue, right and left cheek and  most vulnerable to radiation are soft and hard palate. We note that the ranking of sensitivity of sites is \emph{not} in the same order than if we just consider marginal frequencies of mucositis scores (Figure~\ref{fig:barplot}), due to the confounding of cumulative dose with sites. On the boxplots of cumulative dose by site in Figure~\ref{Oddsfirstobs:fig} we have marked the cumulative dose when mucositis score 3 or 4 first occurred. This shows that for upper and lower lip mucositis occurs rarely and at doses higher than those where it occurs at other sites. We can also see that there is a good overlap in cumulative dose when mucositis first occurred between sites. 

\begin{figure}[!htb]
 \centering
 \includegraphics[scale=0.6]{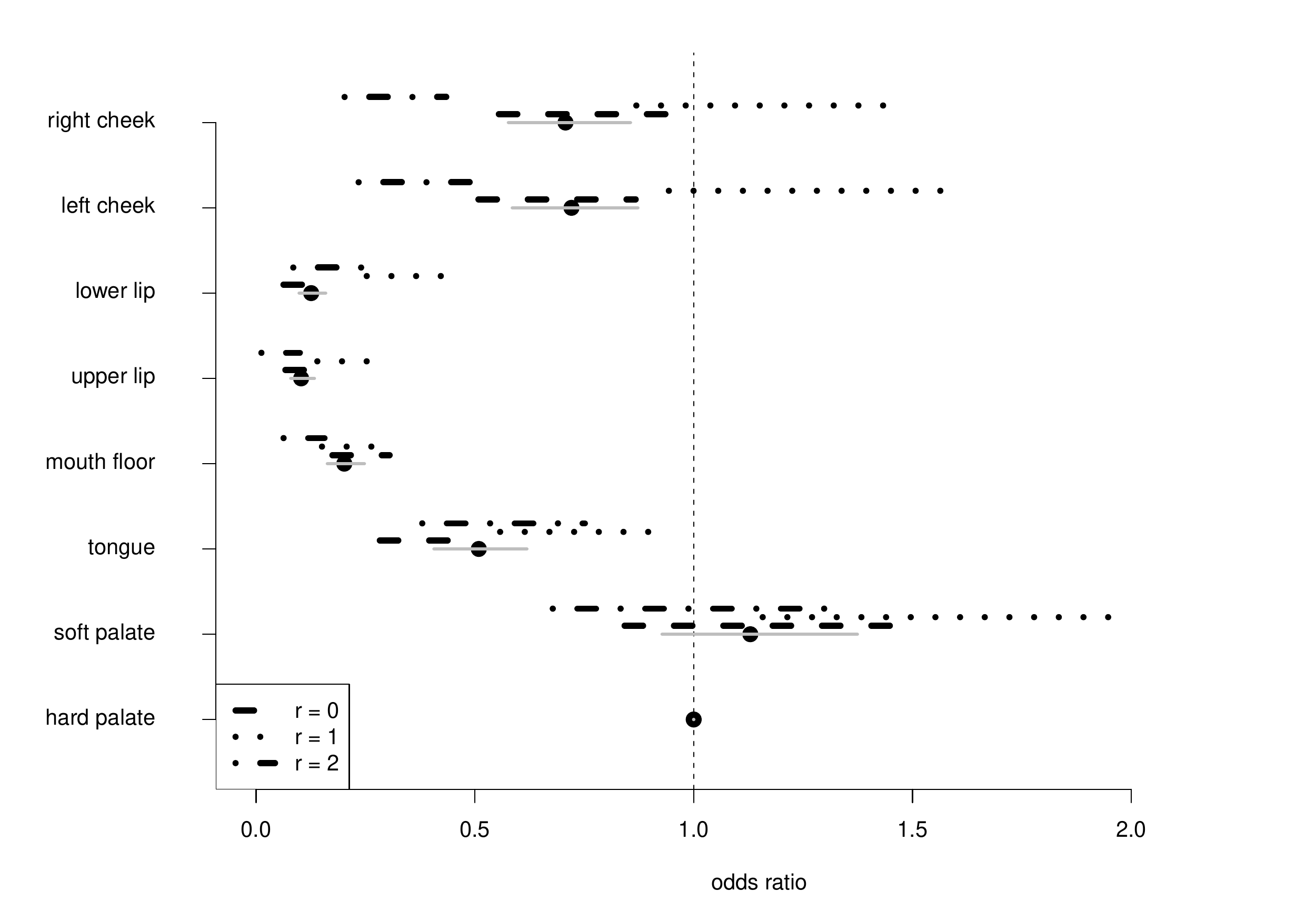}
 \caption{Odds ratios of all the sites with 95\% confidence/credible intervals from the proportional odds model (best model~(\ref{eqn:best})) in gray. Overlaid are the corresponding confidence intervals of odds ratios from logistic regression models with the same terms as the best model~(\ref{eqn:best}) having collapsed the mucositis scores into a binary response using different values of $r= 0, 1, 2$ for the two events $Y \leq r$ and $Y > r$ as shown in Table~\ref{tab:collapse}. Note that we do not use $r=3$,  because in the analysis we combined mucositis scores 3 and 4.}
 \label{propoddsVSlogistic:fig}
\end{figure}

Figure~\ref{propoddsVSlogistic:fig} compares the odds ratios of all sites from the best model~(\ref{eqn:best}) with the odds ratios from logistic regression models with the same terms as the best model~(\ref{eqn:best}), but having collapsed the mucositis scores into a binary response using different values of $r$. For $r = 0$ the results are similar, but the logistic regression model yields wider confidence intervals, for $r = 1$ and $r = 2$ the confidence intervals from the logistic regression deviate substantially, and the answers are not quite as clear cut as the confidence intervals overlap. This demonstrates that we loose efficiency by aggregation. 

Our results depend on the accurate measurement of percentage median dose arriving at individual sites throughout the radiation schedule. It is also worth noting that we have not adjusted for the absolute volume of the sites in our model, but we assume that the volume does not differ substantially between sites. It is future work to investigate whether site volume has an effect on mucositis.  In our analysis we have used cumulative dose rather than biological effective dose, as the cumulative dose achieved the best model fit. 

As in any statistical investigation, inference relies on the validity of the model. We have taken great care to avoid model misspecification by using a flexible multivariate random effects proportional odds model to estimate the effects. The model is complex, but this level of complexity is required in order to reflect the data generating mechanism. We have carried out a thorough model selection based on the AIC, BIC and the root mean square prediction error estimated by 15-fold cross-validation. 
 The model deals with the repeated multivariate ordered categorical outcome measures of individuals via random intercepts for individuals. The model also has shown that the effect of cumulative dose is not linear, and that there is a possible interaction between site and site specific dose, as we could not entirely reject the full model~(\ref{eqn:interaction}).

\section*{Acknowledgements}
This work was supported by funding from the European Community’s Seventh Framework Programme FP7/2011: Marie Curie Initial Training Network MEDIASRES (“Novel Statistical Methodology for Diagnostic/Prognostic and Therapeutic Studies and Systematic Reviews”; www.mediasres-itn.eu) with the Grant Agreement Number 290025.
\bibliographystyle{SageV}


\end{document}